\documentclass[twocolumn]{aastex63} 

\usepackage{xspace}
\usepackage[normalem]{ulem}

\usepackage{amsmath}
\usepackage{rotating}

\graphicspath{{./}{figures/}}

\shorttitle{Multi-messenger Implications of Sub-PeV Diffuse Galactic Gamma-Ray Emission}
\shortauthors{Fang \& Murase}

\begin{document}

\title{Multi-messenger Implications of Sub-PeV Diffuse Galactic Gamma-Ray Emission} 

\author{Ke Fang}
\affil{Department of Physics, Wisconsin IceCube Particle Astrophysics Center, University of Wisconsin, Madison, WI, 53706}
 
\author{Kohta Murase}
\affil{Department of Physics, The Pennsylvania State University, University Park, Pennsylvania 16802, USA}
\affil{Department of Astronomy \& Astrophysics, The Pennsylvania State University, University Park, Pennsylvania 16802, USA}
\affil{Center for Multimessenger Astrophysics, Institute for Gravitation and the Cosmos, The Pennsylvania State University, University Park, Pennsylvania 16802, USA}
\affil{Center for Gravitational Physics, Yukawa Institute for Theoretical Physics, Kyoto, Kyoto 606-8502 Japan}

\begin{abstract}
The diffuse Galactic gamma-ray flux between 0.1 and 1~PeV has recently been measured by the Tibet AS$\gamma$ Collaboration. The flux and spectrum are consistent with the decay of neutral pions from hadronuclear interactions between Galactic cosmic rays and the interstellar medium (ISM). We derive the flux of the Galactic diffuse neutrino emission from the same interaction process that produces the gamma rays.
Our calculation accounts for the effect of gamma-ray attenuation inside the Milky Way and uncertainties due to the spectrum and distribution of cosmic rays, gas density, and infrared emission of the ISM. 
We find that the contribution from the Galactic plane to the all-sky neutrino flux is $\lesssim5-10$\% around 100~TeV. The Galactic and extragalactic neutrino intensities are comparable in the Galactic plane region. Our results are consistent with the upper limit reported by the IceCube and ANTARES Collaborations, and predict that next-generation neutrino experiments may observe the Galactic component. 
We also show that the Tibet AS$\gamma$ data imply either an additional component in the cosmic-ray nucleon spectrum or contribution from discrete sources, including Pevatrons such as superbubbles and hypernova remnants, and PeV electron accelerators. 
Future multi-messenger observations between 1~TeV and 1~PeV are crucial to decomposing the origin of sub-PeV gamma rays. 
\end{abstract}


\section{Introduction}
Diffuse gamma rays with energies between 100~TeV and 1~PeV have recently been detected by the Tibet air-shower gamma (Tibet AS$\gamma$) experiment. The arrival directions of the sub-PeV gamma rays extend over the Galactic plane (GP) and are consistent with being from diffuse cosmic rays in the Galaxy \citep{TibetDiff}. 
The origin of cosmic rays has been an enigma especially around and beyond the knee energy at $E_{\rm knee}\sim3-4$~PeV. Various sources, including superbubbles, hypernovae, supernovae in dense circumstellar material, merger remnants, pulsar wind nebulae, and the Galactic center, have been proposed as potential contributors.  
Dissecting the diffuse gamma-ray emission at MeV-GeV energies~\citep{Hunger:1997we,FermiLAT:2012aa}
to PeV energies is important to understanding the cosmic-ray sources and propagation physics. 

Sub-PeV gamma rays play crucial roles in multi-messenger astrophysics. The detection of $\sim0.1-1$~PeV neutrinos enables us to probe $\sim3-30$~PeV protons beyond the knee energy. The IceCube experiment has measured diffuse TeV-PeV neutrinos that are consistent with an extragalactic origin~\citep{Aartsen:2013bka,Aartsen:2013jdh,Aartsen:2015ita,Aartsen:2020aqd,2019ICRC...36.1017S}. The neutrino and gamma-ray connection has been shown to be powerful for revealing both Galactic and extragalactic sources of neutrinos~\citep{Murase:2013rfa,Ahlers:2013xia,Murase:2015xka}.  

To date diffuse Galactic neutrino emission has not been discovered~\citep{Aartsen_2017,Albert:2018vxw}, despite that it has been predicted to exist for decades~\citep[][for a recent review]{Stecker:1978ah,Kheirandish:2020upj} and studied in light of the IceCube measurements~\citep{Ahlers:2013xia,Anchordoqui:2013qsi,Neronov:2013lza,Joshi:2013aua,Kachelriess:2014oma,Spurio:2014una,2015ApJ...815L..25G,Palladino:2016zoe,Denton:2017csz}. 
In particular, primarily based on previous sub-PeV gamma-ray limits posed by the CASA-MIA~\citep{Borione_1998} and KASCADE experiments, \cite{Ahlers:2013xia} showed that the Galactic contribution to IceCube neutrinos is subdominant~\citep[see also][]{Murase:2015xka}, and the GP may give $\sim3-10$\% of the $10-100$~TeV all-sky neutrino flux with $E_\nu^2\Phi_\nu^{\rm IC}\sim(5-10)\times{10}^{-8}~{\rm GeV}~{\rm cm}^{-2}~{\rm s}^{-1}~{\rm sr}^{-1}$~\citep{Aartsen:2015ita,Aartsen:2020aqd}. This is also consistent with neutrino constraints~\citep[e.g.,][]{Spurio:2014una,Ahlers:2015moa} as well as the latest IceCube and ANTARES results, $\lesssim8.5$~\%~\citep{Albert:2018vxw}. 

The Galactic disk is magnetized, and can be regarded as a cosmic-ray ``reservoir'' with a typical escape time scale of $t_{\rm esc}\sim30~{\rm Myr}~{(R/1~{\rm GV})}^{-\delta}$~\cite[e.g.,][]{Murase:2018utn}, where $R$ is the rigidity and $\delta\sim0.3-0.5$.   
Pionic gamma rays and neutrinos are co-produced when cosmic-ray ions interact with gas and dust particles in the Galaxy~\citep{Hayakawa52,Stecker:1977ua}. 
The detection of diffuse Galactic gamma rays therefore provides a solid reference to the flux level of the diffuse Galactic neutrinos as well as the cosmic-ray confinement and injection around the knee energy. 
In this work, by taking into account details of the gamma-ray attenuation along the line of sight, we evaluate the diffuse Galactic neutrino flux using the sub-PeV gamma-ray flux observed by the Tibet AS$\gamma$ experiment. 
We show that the Galactic contribution can be comparable to the extragalactic flux in the direction to the GP especially below $E_\nu \lesssim 100$~TeV, and may be observed by the next-generation neutrino telescopes. Our conclusion also applies to neutrino emission from unresolved sources in the Galaxy. 

This letter is organized as follows. 
We study the diffuse neutrino emission in $\S\ref{sec:diffuse}$. 
We discuss the potential contribution to sub-PeV gamma rays from discrete sources, including Cygnus Cocoon ($\S\ref{sec:CygnusCocoon}$) and unresolved hadronic ($\S\ref{sec:HNR}$) and leptonic ($\S\ref{sec:leptonic}$) sources. We conclude in $\S\ref{sec:discussion}$.

\section{Multimessenger Connection in Galactic Diffuse Emission}\label{sec:diffuse}
In hadronuclear ($pp$) scenarios, the differential spectrum of neutrinos and gamma rays per interaction are related by \citep[e.g.,][]{Murase:2013rfa,Ahlers:2013xia},
\begin{equation}
E_\nu^2\frac{dN_\nu}{dE_\nu} \approx \frac{3}{2}\left.\left(E_\gamma^2\frac{dN_\gamma}{dE_\gamma}\right)\right|_{E_\nu \approx E_\gamma / 2},
\label{eq:MM}
\end{equation}
considering that the ratio of charged and neutral pions is approximately $2:1$ at high energies and each neutrino carries $\sim1/4$ of the pion energy.  
Neutrinos with characteristic energy carry $\sim3-5$\% of the parent nucleon energy, i.e., $E_\nu\sim(0.03-0.05)E_p$. Equation~(\ref{eq:MM}) is further subject to the gamma-ray attenuation. This is because PeV gamma rays travel only $\sim10$~kpc. In this work, we take into account the effect of the gamma-ray absorption in detail. We do not consider electromagnetic cascades since the effect is small for steep spectra with $\alpha_\nu>2$~\citep[see, e.g.,][]{Murase:2012xs}.

\subsection{Methods}
Below $\vec{x}$ indicates a cylindrical coordinate system with the Galactic center (GC) at the origin and the GP on the $xy$ plane, referred as the GC frame. The projected distance of $\vec{x}$ to the GC on the $xy$ plane is noted as $r$ and the distance to the plane is $z$. This coordinate system is suitable for the description of the diffuse infrared emission and cosmic-ray source distribution in the Galaxy, which are approximately cylindrically symmetric. For observation of neutrinos and gamma rays at the solar neighborhood we will use the Galactic coordinate $\vec{x}_g$. Taking the direction toward the Galactic Center as the $x$-axis, we have $\vec{x}_g = (s\cos b\,\cos l, s\cos b\,\sin l,s\,\sin b)$, where $s$ is the distance to the observer (the Sun), $l$ and $b$ are the Galactic longitude and latitude, respectively. $\vec{x}_g$ and $\vec{x}$ are converted by $\vec{x} = \vec{x}_g + \vec{x}_{\rm obs}$, where $\vec{x}_{\rm obs}$ is the coordinate of the observer in the GC frame. 

The inverse of the mean free path for a gamma ray of energy $E_\gamma$ and direction $\hat{u}$ at a space point $\vec{x}$ is,
\begin{equation}
\lambda_{\gamma\gamma}^{-1}(E_\gamma, \hat{u}, \vec{x}) = \int d\Omega   (1 - \hat{u}\cdot\hat{k}) \int d\varepsilon \frac{dn}{d\varepsilon d\Omega} (\vec{x}) \sigma_{\gamma\gamma}(E_\gamma \varepsilon(1-\hat{u}\cdot\hat{k})).  
\end{equation}
In the equation $\sigma_{\gamma\gamma}$ is the cross section for pair production $\gamma\gamma\rightarrow e^+e^-$,
Here $dn/d\varepsilon d\Omega(\vec{x})$ is the number density of target photons per unit energy per unit solid angle at the position $\vec{x}$ and direction $\hat{k}$. For photons above 10~TeV, the main target photons are the cosmic microwave background (CMB) and the infrared emission by dust. The former is uniform within the Galaxy while the latter depends on $r$ and $z$. We describe the calculation of $dn/d\varepsilon d\Omega(\vec{x})$ and discuss the attenuation by different radiation fields in Appendix~\ref{Appendix:IR}. 

The ${\gamma\gamma}$ optical depth $\tau_{\gamma\gamma}$ for a photon traveling from an initial position $\vec{x}_0$ to an observer at $\vec{x}_{\rm ob}$ is 
\begin{equation}
\tau_{\gamma\gamma}(E_\gamma, \vec{x}_0, \vec{x}_{\rm ob}) = \int_0^{|\vec{x}_{\rm ob}-\vec{x}_0|} ds~ \lambda_{\gamma\gamma}^{-1}(E_\gamma, \hat{u}, \vec{x}_0 + s \hat{u})
\end{equation}
with $\hat{u} = (\vec{x}_{\rm ob} - \vec{x}_0) / |\vec{x}_{\rm ob} - \vec{x}_0|$. 
The probability for a photon to survive from the pair production is~\citep{Vernetto_Lipari16},
\begin{equation}
P_{\gamma, \rm surv}(E_\gamma, \vec{x}_0,  \vec{x}_{\rm ob}) = \exp\left(-\tau_{\gamma\gamma}(E_\gamma, \vec{x}_0,\vec{x}_{\rm ob})\right)
\end{equation}

The averaged gamma-ray intensity from a region of solid angle $\Delta\Omega$ is the sum of the photons that have survived from all the sources in that area,
\begin{eqnarray}
E_\gamma^2 \Phi_\gamma^\Omega 
&\approx&\frac{1}{\Delta\Omega} \int d\Omega~\int ds\,n_{\rm CR}n_{N}(\vec{x}_0) \sigma_{pp} c  
 \\ \nonumber &\times& 
\left(E_\gamma^2\frac{dN_\gamma}{dE_\gamma}\right)  \frac{1}{4\pi |\vec{x}_0 - \vec{x}_{\rm ob}|^2} P_{\gamma, \rm surv}(E_\gamma, \vec{x}_0,\vec{x}_{\rm ob})
\end{eqnarray}
where $n_{\rm CR}n_{N}(\vec{x}_0)$ is the product of the CR and gas/molecular densities. $\sigma_{pp}$ is the inelastic $pp$ cross section, which moderately increases from $\sim 40$~mb at $E_p=100$~TeV to $\sim 70$~mb at $E_p=10$~PeV~\citep[e.g.,][]{PhysRevD.98.030001}. This energy dependence impacts the shapes of the intrinsic gamma-ray spectrum and the neutrino spectrum in the same way.


The all-flavor neutrino flux can be computed in the same way, 
\begin{eqnarray}
E_\nu^2 \Phi_\nu^\Omega \approx\frac{1}{\Delta\Omega} \int d\Omega\int ds\frac{n_{\rm CR}n_{N}(\vec{x}_0) \sigma_{pp} c}{4\pi |\vec{x}_0 - \vec{x}_{\rm ob}|^2}
\left(E_\nu^2\frac{dN_\nu}{dE_\nu}\right).   
\end{eqnarray}
Notice that the chance of a neutrino interaction inside the Galaxy is negligible. In other words, the survival probability of high-energy neutrinos is always 1,

Using equation~(\ref{eq:MM}), we can write the neutrino intensity using the gamma-ray intensity in the Galactic coordinate, 
\begin{eqnarray}
\label{eqn:Fnu_Fg}
&& E_\nu^2 \Phi_\nu^\Omega \approx \frac{3}{2} \left.\left(E_\gamma^2 \Phi_\gamma^\Omega\right)\right|_{E_\gamma = 2E_\nu} \\ \nonumber
&\times&\frac{\int ds \int\cos{b}\, db\int dl \, n_{\rm CR}n_{N}(s, b, l)}{ \int ds \int\cos{b}\,db \int dl \, n_{\rm CR}n_{N}(s, b, l) P_{\gamma,\rm surv}(E_\gamma = 2E_\nu, s, b, l)},
\end{eqnarray}
which is improved compared to equation~2 of \cite{Ahlers:2013xia}. 
 
To account for the uncertainty caused by the $n_{\rm CR}n_{N}$ distribution, which depends on details of the CR propagation and gas density distribution, we consider the two limits. In the first model, as in the leaky box model, cosmic rays are assumed to be uniformly distributed within the disk (that is smaller than the cosmic-ray halo),
\begin{equation}\label{eqn:uniform}
n_{\rm CR}n_{N}(r, z)\propto
\begin{cases} 1 & r < R_{\rm disk}\, \mbox{and}\, |z|<z_{\rm disk}   \\ 
0 & \mbox{otherwise }
\end{cases},
\end{equation}
with $R_{\rm disk}=15$~kpc and $z_{\rm disk}=0.2$~kpc (that is compatible with the scale height of HI gas). 
This gives a conservative estimate on the Galactic diffuse emission~\citep[see, e.g.,][]{Ahlers:2013xia,Ahlers:2015moa}. 
In the second model, we assume that it follows the spatial distribution of supernova remnants (SNRs),
\begin{equation}\label{eqn:SNR}
n_{\rm CR}n_{N}(r, z)\propto \left(\frac{r}{R_\odot}\right)^\zeta\exp{\left[-\eta\left(\frac{r-R_\odot}{R_\odot}\right)-\frac{|z|}{z_g}\right]}.
\end{equation}
where $R_\odot=8.5$~kpc is the solar distance from the GC and  the following parameter values are adopted,   $\zeta=1.09$, $\eta = 3.87$ \citep{2015MNRAS.454.1517G} and $z_g=0.2$~kpc \citep{2012JCAP...01..011B}. This model is appropriate when we consider GP emission from a number of discrete sources. But this would give a optimistic estimate on the Galactic diffuse emission since cosmic rays diffuse out from the sources.   

Finally, the total neutrino (or gamma-ray) flux from the region of interest with $\Omega$ can be derived by the flux from the observed sky region, $\Omega_{\rm obs}$, through  
\begin{eqnarray}\label{eqn:Fi_Fj}
E_i^2 F_{i,\rm GP} &=& E_i^2 F_{i,\rm obs} \nonumber\\
&\times&\frac{\int ds \int^\Omega d\Omega\, n_{\rm CR}n_N(s, b, l)P_{i, \rm surv}(s, b, l)}{\int ds \int^{\Omega_{\rm obs}}d\Omega \, n_{\rm CR}n_N(s, b, l) P_{i,\rm surv}(s, b, l)}\,\,\,\,\,\\
&=&\frac{\Delta\Omega n_{\rm CR}^{\Omega}\tau_{pp}^{\Omega} \tilde{P}_{i,\rm surv}^{\Omega}}{\Delta\Omega_{\rm obs} n_{\rm CR}^{\Omega_{\rm obs}}\tau_{pp}^{\Omega_{\rm obs}} \tilde{P}_{i,\rm surv}^{\Omega_{\rm obs}}}
\end{eqnarray}
where $i$ indicates either $\nu$ or $\gamma$, $n_{\rm CR}^{\Omega}$ is the averaged CR density and $\tau_{pp}^\Omega$ is the averaged $pp$ optical depth~\citep{Ahlers:2013xia}. 
 
\begin{figure}[t]
\centering
\includegraphics[width=.49\textwidth]{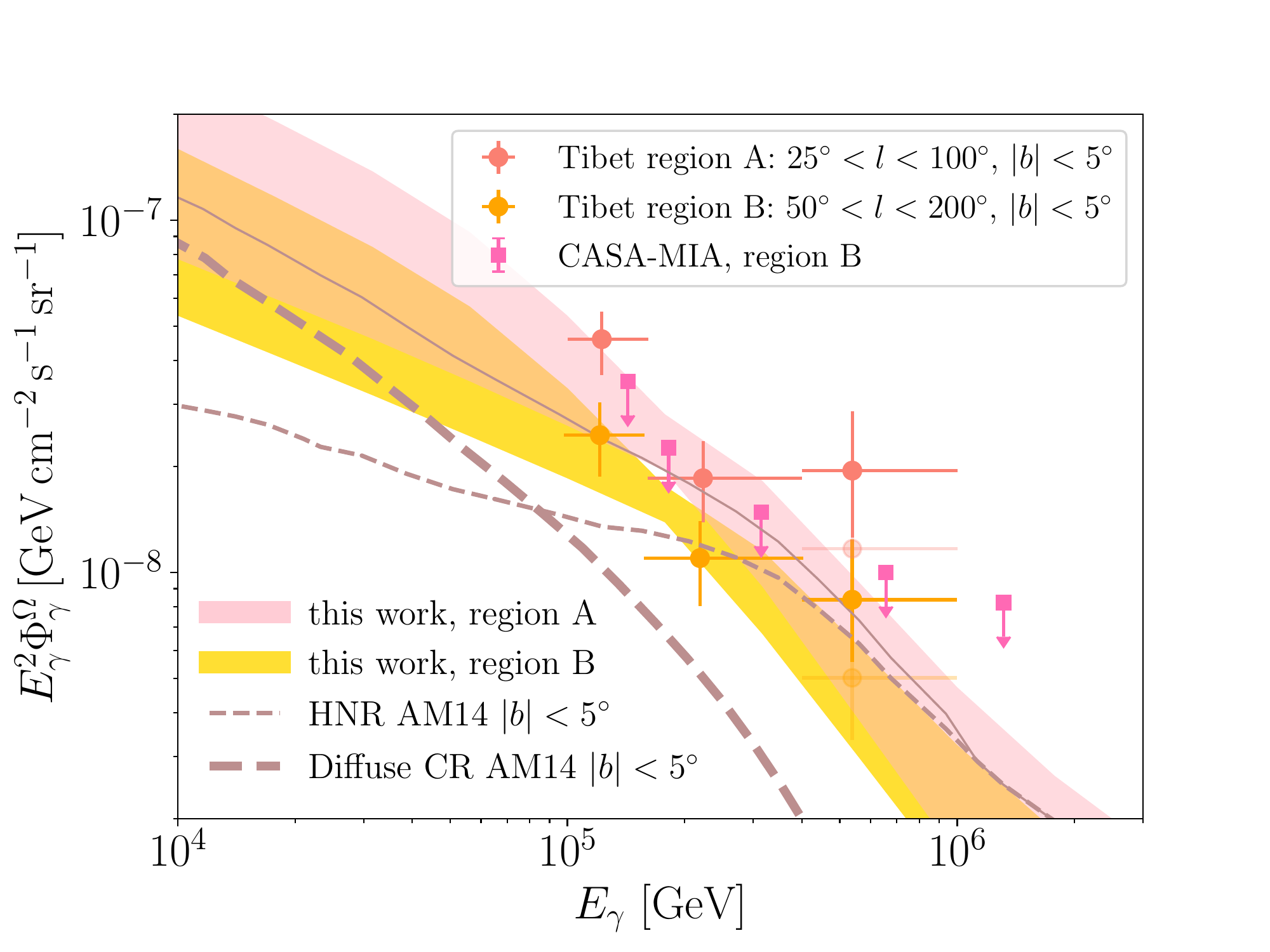}
\caption{\label{fig:Flux}The diffuse Galactic gamma-ray intensity from two sky regions, {\it region A}: $25^\circ < l < 100^\circ$, $|b|<5^\circ$, and {\it region B}: $50^\circ < l < 200^\circ$, $|b|<5^\circ$. 
The red and orange data points are the Tibet AS$\gamma$ measurement of the diffuse $\gamma$-ray emission from the two regions \citep{TibetDiff}. In the last energy bin, the fainter data points indicate the residual intensity after removing events relevant to Cygnus Cocoon. The red and orange bands are the best-fit $\gamma$-ray models derived in this work, accounting for uncertainties in the gamma-ray attenuation and cosmic-ray models. 
The brown long and short dashed curves indicate the diffuse gamma-ray spectra for the GP and unresolved hypernova remnants, respectively, which are taken from \citet{Ahlers:2013xia} for $|b|<5^\circ$. The thin solid curve shows the sum of the two components, which demonstrates that the Tibet AS$\gamma$ data are also consistent with a prediction with source emission. 
}
\end{figure}

\begin{figure}[htp]
\centering
\includegraphics[width=.49\textwidth]{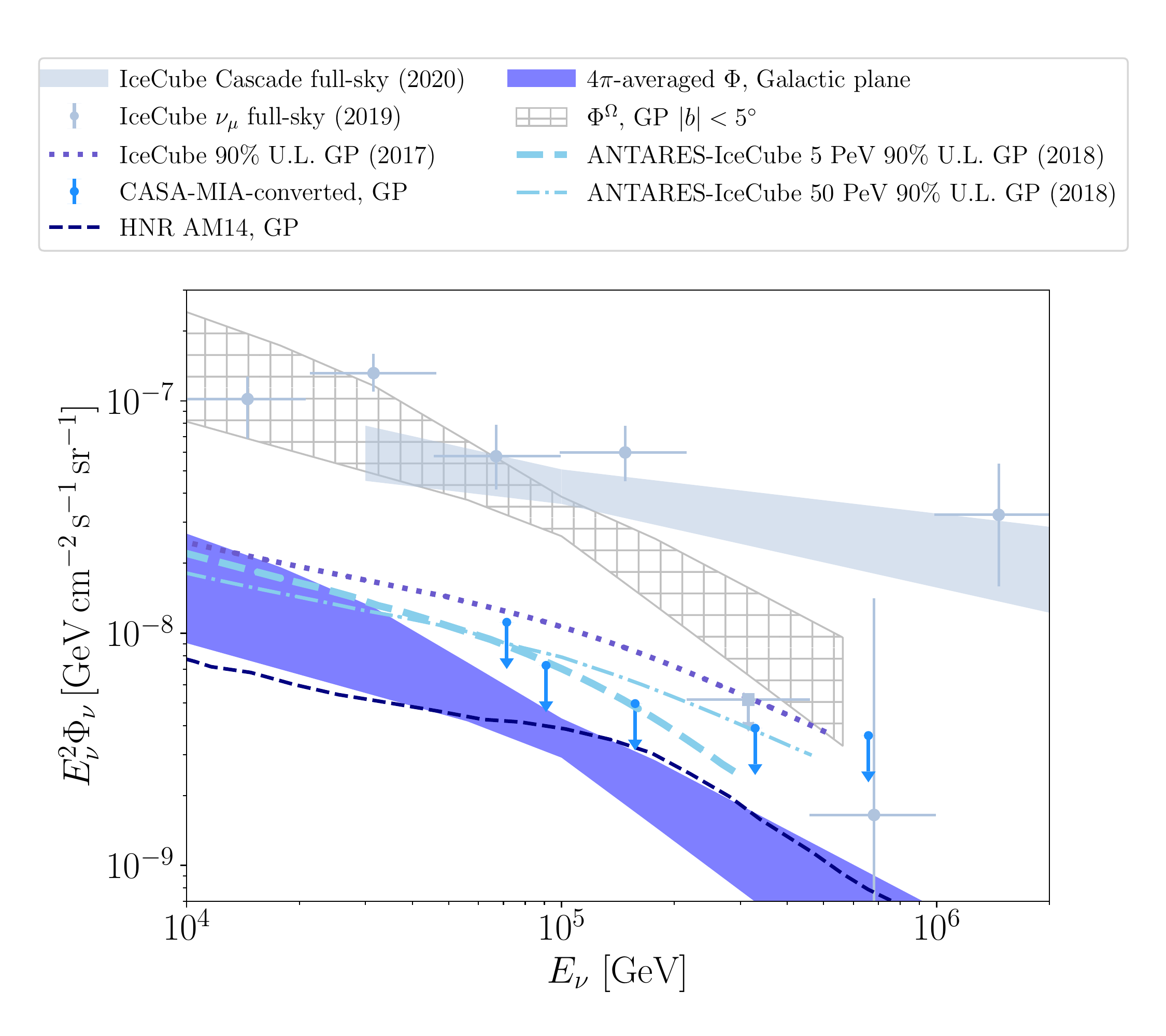}
\caption{\label{fig:fullSkyAveragedPhi} 
All-sky-averaged intensity of all flavor diffuse neutrinos from the GP, compared to neutrino observations. 
The GP neutrino intensity, $E_\nu^2\Phi_\nu$, (blue shaded band) is derived with the best-fit gamma-ray intensities in Figure~\ref{fig:Flux}.  The model is consistent with the combined upper limits at 90\% confidence level posed by ANTARES and IceCube (sky blue dashed and dash-dotted curves; \citealp{2018ApJ...868L..20A}), the 90\% limits with 7-year IceCube data (blue dotted curve; \citealp{Aartsen_2017}), and the upper limits on neutrinos from the GP (blue downward arrows), which are derived from the CASA-MIA gamma-ray limits in region B, assuming that sources follow the SNR distribution (cyan downward arrows; \citealp{Borione_1998}). The hatched band shows the intensity $E_\nu^2\Phi^\Omega_\nu$ of the $|b|<5^\circ$ region, which is comparable to the isotropic neutrino background from the IceCube Cascade (light blue data points; \citealp{Aartsen:2020aqd}) and muon neutrino (light blue shaded area; \citealp{2019ICRC...36.1017S}) data below $\sim 100$~TeV.  
}
\end{figure}

\subsection{Results}\label{sec:results}
Two neutrino spectral models are used to account for the uncertainty in the cosmic-ray nucleon spectrum (see Appendix~\ref{appendix:nuSpectra}). 
The gamma-ray intensity $\Phi_\gamma^\Omega$ from a sky region of solid angle $\Omega$ is obtained from equations~(\ref{eqn:Fnu_Fg}, \ref{eqn:Gaisser},  \ref{eqn:break}) and normalized by fitting to the Tibet AS$\gamma$ data using a $\chi^2$ statistic,
\begin{eqnarray}\label{eqn:chi2}
\chi^2 = \sum_{i, j} \frac{(\Phi^{\Omega_j}_{\gamma,\rm obs}(E_i) - \Phi_{\gamma}^{\Omega_j}(E_i))^2}{\left(\sigma^{\Omega_j}_{\gamma, \rm obs}(E_i)\right)^2}.
\end{eqnarray}
In this equation, $\Phi^{\Omega_j}_{\gamma,\rm obs}(E_i)$ and $\sigma^{\Omega_j}_{\gamma,\rm obs}(E_i)$ are the observed intensity and uncertainty of diffuse gamma rays from sky region $j$ in energy bin $i$, respectively. Two sky regions are considered, namely, {\it region A} with $25^\circ < l<100^\circ$ and $|b|<5^\circ$, and {\it region B} with $50^\circ < l < 200^\circ$ and $|b|<5^\circ$. In each energy bin, the upper (lower) error is used if the model is above (below) the mean of the measurement. \citet{TibetDiff} notes that $40\%$ events above 398~TeV in each of the two sky regions are close to the Cygnus Cocoon (see our discussion in Section~\ref{sec:CygnusCocoon}). Therefore, our fit in the highest energy bin uses $60\%$ of the measured values. The fit uses six flux points from Tibet AS$\gamma$ and one free parameter, the flux norm of equation~\ref{eqn:Gaisser} or \ref{eqn:uniform}, and thus has a total of five degrees of freedom. The resulted chi-square per degree of freedom is $\chi^2\sim1.6-2.0$ and $\chi^2 \sim 0.7-1.0$ for the uniform and SNR source distribution models, respectively.  

The best-fit gamma-ray intensities for the two sky regions measured by Tibet AS$\gamma$ are shown as colored bands in Figure~\ref{fig:Flux}. The boundaries of the bands are decided by the minimum and maximum values from the four cases in use, which include the two gamma-ray emissivity distribution models and the two cosmic-ray nucleon spectra. In particular, in these shaded bands, upper boundaries below $\sim 150$~TeV are determined by the nucleon spectrum Model A (equation~\ref{eqn:Gaisser}), whereas those above are governed by the nucleon spectrum Model B (equation~\ref{eqn:break}). See also Appendix~\ref{appendix:nuSpectra}.
We find that the sub-PeV gamma-ray spectral shape is barely affected by uncertainties in the gamma-ray attenuation. Rather, it depends on the shape of the cosmic-ray nucleon spectrum.
For example, if the cosmic-ray nucleon spectrum has a break at $E_*=0.9$~PeV~\citep{Gaisser:2013bla}, the gamma-ray spectrum should be steepened at $\sim0.08E_*\sim80$~TeV, which causes a tension with the Tibet AS$\gamma$ data at the highest-energy bin. The tension can readily alleviated if discrete sources make a significant contribution at sub-PeV energies as we discuss in Section~\ref{sec:source}. 

Now, let us consider the GP contribution to IceCube neutrinos. 
Figure~\ref{fig:fullSkyAveragedPhi} shows the $4\pi$-averaged intensity of diffuse Galactic neutrinos (shaded blue band) calculated using the best-fit gamma-ray models in Figure~\ref{fig:Flux}, equations~(\ref{eqn:Fi_Fj}) and (\ref{eqn:Fnu_Fg}). 
The GP contribution to the all-sky neutrino flux depends on neutrino energy, and is found to be $\lesssim5-10$\% in the 100~TeV range. We caution that the exact value depends on the all-sky IceCube flux and varies with energy depending on the spectra. 
Analytically, noting that the all-sky-averaged intensity is denoted as $\Phi_\nu\equiv(\int d\Omega~\Phi_\nu^{\Omega}/4\pi)=(\Delta\Omega_{\rm GP}/4\pi)\Phi_\nu^{\rm GP}$, we may write~
\begin{eqnarray}
\frac{\Delta\Omega_{\rm GP} E_\nu^2\Phi_\nu^{\rm GP}}{4\pi  E_\nu^2\Phi_\nu^{\rm IC}}
\sim5\%{\left(\frac{E_\nu^2\Phi_\nu^{\rm IC}}{5\times{10}^{-8}~{\rm GeV}{\rm cm}^{-2}~{\rm s}^{-1}~{\rm sr}^{-1}}\right)}^{-1}\nonumber\\
 \times
\left(\frac{\Delta\Omega_{\rm GP}}{1.1~\rm sr}\right)\left(\frac{E_\gamma^2\Phi_\gamma^{\rm GP}|_{E_\gamma=2E_\nu}}{2\times{10}^{-8}~{\rm GeV}{\rm cm}^{-2}~{\rm s}^{-1}~{\rm sr}^{-1}}\right). \,\,\,\,\,\,\,\,\,\,
\end{eqnarray}
Here the effective solid angle of the GP is assumed to be $\Delta\Omega_{\rm GP}=2\sin b\Delta l\simeq1.1~{\rm sr}~{(|b|/5^\circ}){(\Delta l/360^\circ)}$ and $E_\gamma^2\Phi_\gamma^{\rm GP}\sim E_\gamma^2\Phi_\gamma^{\Omega_{\rm GP}}$~\footnote{The all-sky estimate may depend on the distribution in the vertical direction of the disk. For a a scale height of $0.1$~kpc, we have $E_\gamma^2\Phi_\gamma^{\rm GP}/E_\gamma^2\Phi_\gamma^{\Omega_{\rm GP}}\simeq1.6$~\citep{Ahlers:2013xia}.}. 
This value is also consistent with previous results~\citep{Ahlers:2013xia,Ahlers:2015moa}. Following \cite{Ahlers:2013xia}, we also show limits on GP neutrino emission by converting the CASA-MIA upper limits through equations~\ref{eqn:Fnu_Fg} and \ref{eqn:Fi_Fj}, assuming that sources follow SNR distribution. 
As clearly seen in Figure~\ref{fig:fullSkyAveragedPhi}, our results obtained with the new Tibet AS$\gamma$ data are also consistent with the upper limits posed by \citet{2018ApJ...868L..20A} and \citet{Aartsen_2017} through independent neutrino observations. This is not surprising because IceCube neutrinos are consistent with an isotropic distribution, and they mostly come from the region outside the GP~\cite[][for a review]{Kheirandish:2020upj}. 
 
The hatched gray band in Figure~\ref{fig:fullSkyAveragedPhi} shows the neutrino intensity of the GP with $|b|<5^\circ$. The Tibet AS$\gamma$ data imply that the intensity of diffuse Galactic neutrinos can be comparable to the isotropic diffuse neutrino background in the GP region in the $\sim 10-100$~TeV range, i.e., 
\begin{equation}
E_\nu^2\Phi_\nu^{{\Omega}_{\rm GP}}\sim E_\nu^2\Phi_\nu^{\rm IC}.
\end{equation}
Our result suggests that the diffuse Galactic emission below $\sim 100$~TeV can be promisingly observed by next-generation neutrino telescopes such as IceCube-Gen2 \citep{2020arXiv200804323T}, KM3Net \citep{2016JPhG...43h4001A}, and Baikal-GVD \citep{2018arXiv180810353B}.
At higher energies, the source contribution can be relevant (see the HNR curve in Figure~\ref{fig:fullSkyAveragedPhi} obtained by averaging the flux within $b=10^\circ$), which is also encouraging for next-generation high-energy neutrino telescopes. Even non-detections would also be useful, because as discussed in the next section, the diffuse gamma rays may potentially come from discrete sources. 

\section{Contribution of Discrete sources to Sub-PeV Gamma Rays and Neutrinos}\label{sec:source}
The sub-PeV diffuse emission discussed in Section~\ref{sec:diffuse} comes from ions that were injected into the Galaxy by sources $\sim10^4-10^6$~yr ago, propagate in the Galactic magnetic field, and interact with gas and dust in the Milky Way. In this section we focus on the contribution of discrete sources to the sub-PeV gamma rays observed by the Tibet AS$\gamma$ experiment. 

Sub-PeV gamma rays from discrete sources are naturally expected for two main reasons.
First, as noted in Section~\ref{sec:diffuse}, a nucleon spectrum with a break energy lower than $\sim 1$~PeV, which is suggested by the composition modeling around the knee, is in tension with the last Tibet AS$\gamma$ data point. 
Second, theoretically, it is natural that gamma rays are produced inside or in the vicinity of Pevatrons, including the recently detected TeV source Cygnus Cocoon and other unresolved hadronic and leptonic source populations. Harder spectra of source neutrinos and gamma rays may dominate over the steep spectra of diffuse secondaries at the highest energies. 

\subsection{Hadronic Sources}
\subsubsection{Cygnus Cocoon}\label{sec:CygnusCocoon}
\begin{figure}[t]
\centering
\includegraphics[width=.49\textwidth]{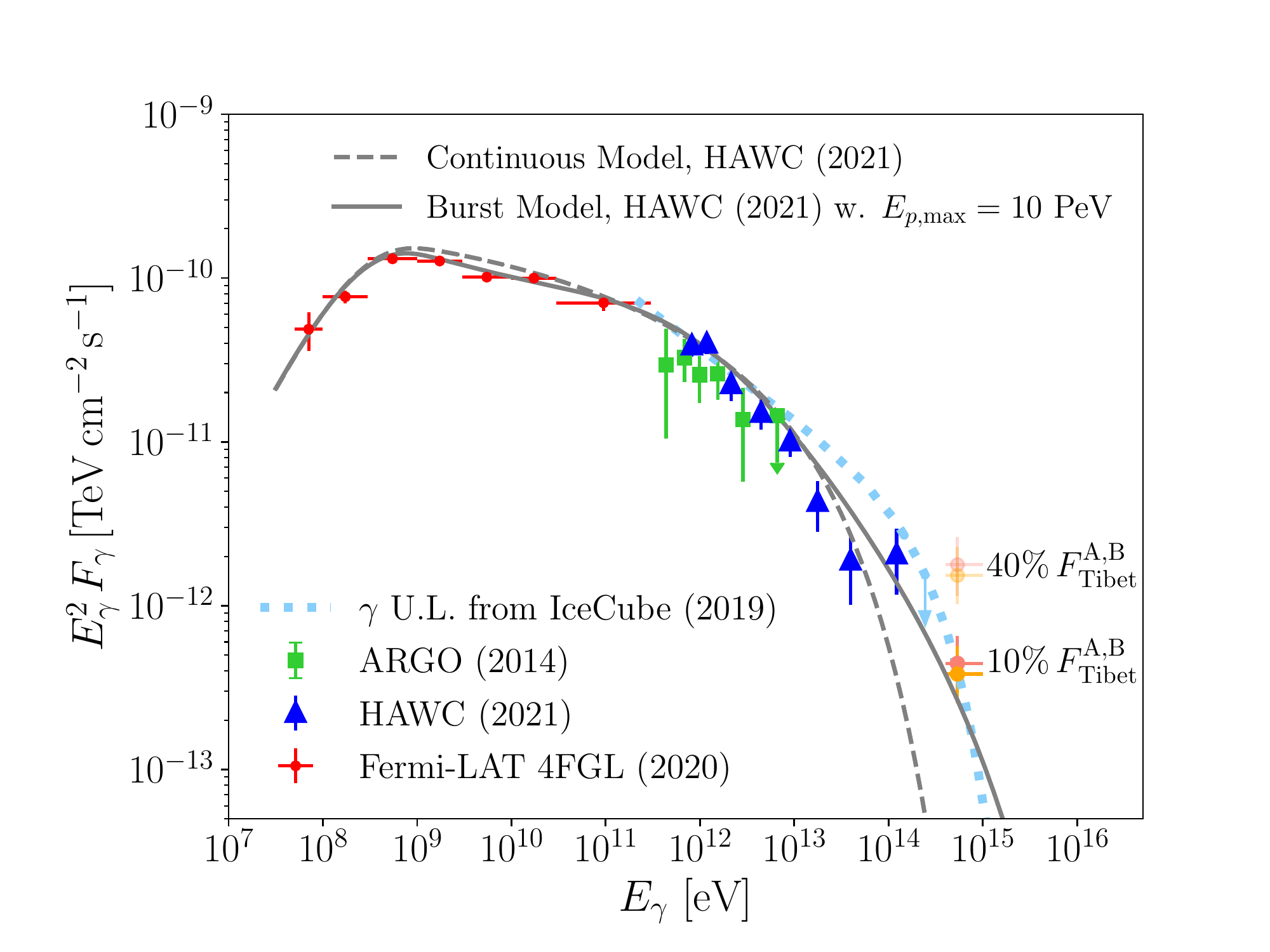}
\caption{\label{fig:Cocoon}
Spectral energy distribution of the Cygnus Cocoon measured by {\it Fermi}-LAT \citep{2020ApJS..247...33A}, ARGO-YBJ \citep{2014ApJ...790..152B}, and HAWC \citep{2021NatAs.tmp...50A}. The light pink and orange flux points indicate 40\% of the Tibet AS$\gamma$ flux of regions A and B \citep{TibetDiff}. The thick pink and orange markers additionally scale the fluxes to the HAWC size of the Cygnus Cocoon. The blue dotted curve shows the limit on the $\gamma$-ray flux based on the non-detection of neutrinos from the region by IceCube \citep{2019ICRC...36..932K}. The two $\gamma$-ray emission models from \citet{2021NatAs.tmp...50A} are shown for comparison. A significant detection of the Cygnus Cocoon at the estimated flux level may favor the burst model and the presence of a Pevatron.  
}
\end{figure}

\citet{TibetDiff} indicates that $\sim 40\%$ of the events in their highest-energy bin are detected within $4^\circ$ around the Cygnus Cocoon. Extended $1-200$~TeV gamma-ray emission from the Cygnus Cocoon has recently been reported by the HAWC Observatory, with emission above 100~TeV detected at $\sim2.4\sigma$ significance level \citep{2021NatAs.tmp...50A}. 
The gamma-ray spectrum can be explained by protons that either have been continuously injected over the lifetime of the stellar clusters (a few Myr), or were produced by a recent (sub-Myr) burst-like event. The latter scenario invokes the presence of a Pevatron, specifically, a PeV proton accelerator with a hard spectrum $dN_p/dE_p \propto E_p^{-2.1}$. 

The Cygnus OB2 association~\citep[e.g.,][]{2020NewAR..9001549W} has been among the most promising sites for cosmic-ray acceleration for many years~\citep[see][for a review]{Bykov:2014asa}. Superbubbles may accelerate cosmic rays up to PeV energies or beyond via multiple shocks and turbulence~\citep{1992MNRAS.255..269B,2000APh....13..161K}. 
It has been shown that cosmic rays escaping from star clusters or superbubbles may explain cosmic rays above the knee energy~\citep{Murase:2018utn,Zhang:2019sio}. 
The Cygnus region has also been of much interest as the promising source of high-energy neutrinos~\citep[e.g.,][]{Anchordoqui:2006pe,Beacom:2007yu,Halzen:2007ah,Halzen:2016seh}. 

Figure~\ref{fig:Cocoon} suggests that observations of the Cygnus Cocoon region above 200~TeV may disentangle the continuous and burst scenarios of \citet{2021NatAs.tmp...50A}. The light pink and orange data points correspond to 40\% of the Tibet AS$\gamma$ flux above 398~TeV from regions A and B. Since the radius of the Cocoon is measured to be $\sim2^\circ$ at $1-100$~TeV by HAWC, the thick pink and orange markers show a more conservative estimation of the Tibet AS$\gamma$ flux using the average event number within the HAWC radius. We caveat that the pink and orange flux points in Figure~\ref{fig:Cocoon} are approximate. The actual flux depends on the $\gamma$-ray morphology and the detector exposure. No high-energy neutrino emission has been detected from the Cygnus Cocoon. The blue dotted curve shows the gamma-ray upper limit converted from the IceCube limit on this source~\citep{2019ICRC...36..932K}. 
For comparison, we overlay the continuous model and the burst model from \citet{2021NatAs.tmp...50A}. In particular, we here update the maximum proton energy in the burst model from 2~PeV to 10~PeV to accommodate the estimated Tibet AS$\gamma$ flux. 
The other model parameters remain the same. 
We find that the burst scenario, hence a Pevatron, would be favored if the flux above 400~TeV reaches $\gtrsim 3\times 10^{-13}\,\rm TeV\,cm^{-2}\,s^{-1}$ assuming $\sim$30\% measurement uncertainty.

\subsubsection{Hypernova Remnants}\label{sec:HNR}
Recent optical observations have revealed that energetic supernovae with a kinetic energy of ${\mathcal E}_{\rm ej}\gtrsim10^{52}$~erg are not negligible as the cosmic-ray energy budget~\citep[e.g.,][]{Murase:2018utn}, and their rate is about $\sim3$\% of the core-collapse supernova rate that is $\sim3$ per century. Energetic supernovae so-called hypernovae (that are mostly broad-line Type Ibc supernovae) and trans-relativistic supernovae, which are often associated with low-luminosity gamma-ray bursts, have been widely discussed as cosmic-ray accelerators, and they may accelerate cosmic rays up to $\sim10-100$~PeV energies~\citep[e.g.,][]{Sveshnikova:2003sa,Murase:2013rfa,Senno:2015tra}. 
It has been argued that the X-ray emission from the Cygnus region can be attributed to a hypernova remnant~\citep{2013PASJ...65...14K,Bluem:2021zas}. The burst model discussed in the previous subsection is consistent with such a model. 
The required cosmic-ray input, $\sim10^{51}$~erg, is comparable to the energy amount of cosmic rays accelerated by a hypernova. Dozens of hypernova remnants are expected to exist in the Milky Way, and a fraction of IceCube neutrinos may come from them~\citep{Fox:2013oza}. There may be a few hypernova remnants in region A and B, and one of them could be in the Cygnus region. 
As shown in Figure~\ref{fig:Flux}, the model~\citep{Ahlers:2013xia} is consistent with the Tibet AS$\gamma$ data. This demonstrates the potential relevance of contributions from discrete sources, and we stress that other candidate sources are also possible.

\subsection{Leptonic Sources}\label{sec:leptonic}
\begin{figure}[t]
\centering
\includegraphics[width=.49\textwidth]{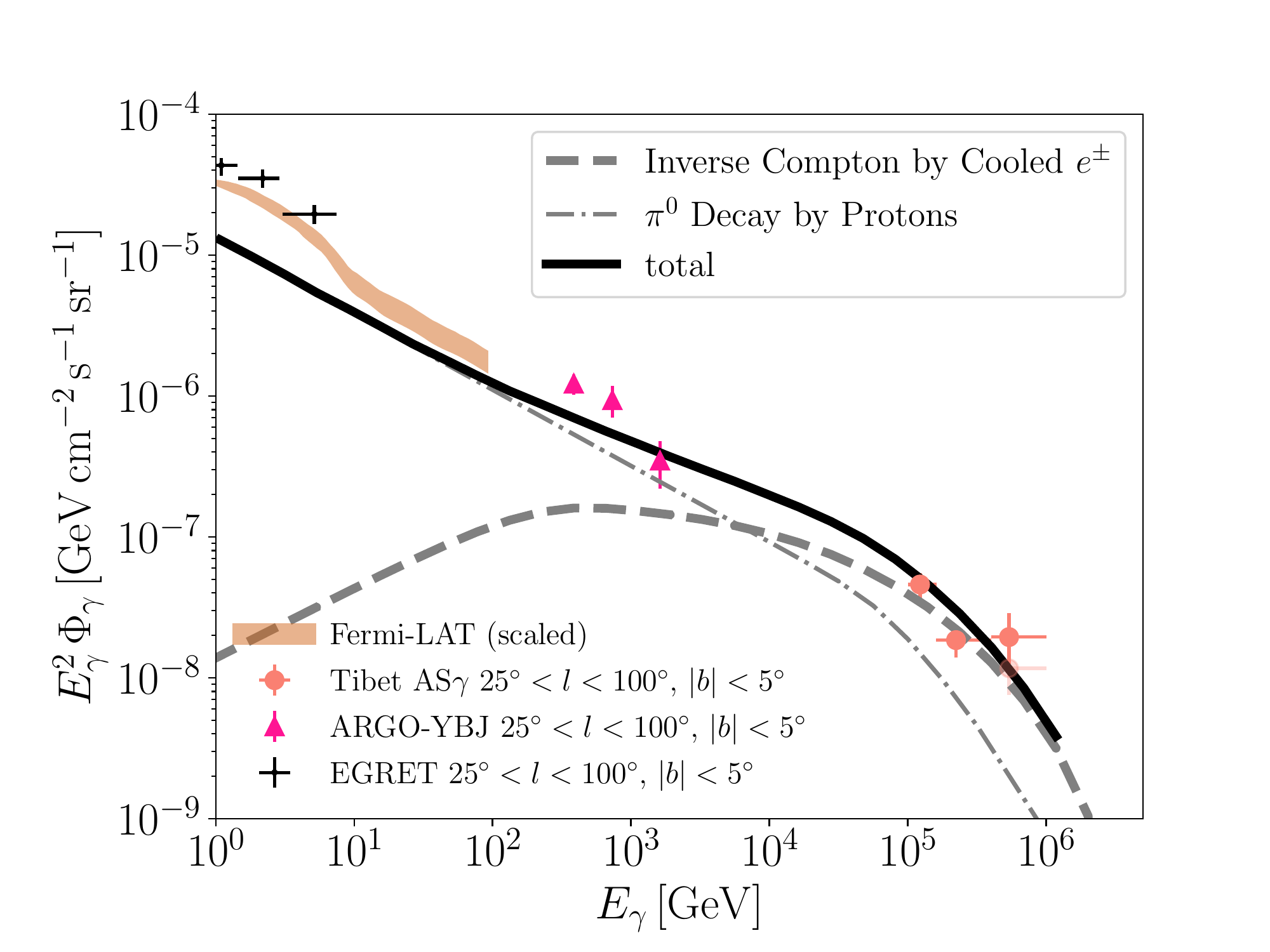}
\caption{\label{fig:Leptonic}
Demonstration of a hybrid $\gamma$-ray emission model, in which the inverse Compton of relativistic electrons (grey dashed curve) explains the Tibet AS$\gamma$ measurement in the region $25^\circ < l<100^\circ$ (red round data points), and $\pi^0$ decay by Galactic diffuse protons (grey dash-dotted curve) explains the lower-energy observations of the same region by EGRET (black plus markers; \citealp{1997ApJ...481..205H}), {\it Fermi}-LAT (brown shaded region, scaled from \citealp{Ackermann_2012} to the EGRET flux), and ARGO-YBJ (pink triangle data points; \citealp{Bartoli_2015}). The electrons are assumed to have an intrinsic spectrum $dN/dE_e\propto E_e^{-2}$ and maximum energy $E_{e,\rm max}=3$~PeV. 
}
\end{figure}

While the sub-PeV $\gamma$-ray emission can be plausibly explained by the decay of neutral pions from hadronuclear interactions~\citep{TibetDiff}, a leptonic origin may not yet be excluded.
Figure~\ref{fig:Leptonic} demonstrates such a scenario, where the Tibet AS$\gamma$ data can in principle explained by electrons that upscatter the CMB. We here assume that relativistic electrons are injected by discrete Galactic sources such as pulsar wind nebulae, confined close to the vicinity of the emission region, while being cooled via synchrotron radiation in the Galactic magnetic field and inverse-Compton scattering with the CMB. The steady-state electron distribution is calculated by solving the transport equation, with $B=3~\mu$G and $u_{\rm CMB} = 0.26\,\rm eV\,cm^{-3}$ for the energy density of magnetic field and the CMB. We note that $B$ near the sources could be higher than the average ISM field strength that we take. Besides, our example model does not account for the IR radiation field at the emission sites which could further contribute to gamma-ray production below $\sim 10$~TeV.
At tens to hundreds of TeV electron energies the cooling timescale is much shorter than the diffusion timescale, therefore the diffusion process is negligible for the calculation. Assuming that $Q_e\propto E_e^{-2}$ up to a maximum energy $E_{e,\rm max}=3~\,\rm PeV$ and source emission time $\sim 0.1$~Myr (just for demonstrative purposes), we find that the total electron power $L_e=\int_{m_ec^2}^{E_{e,\rm max}}\sim 10^{37}\, (d/5\,{\rm kpc})^{2}\,\rm erg\,s^{-1}$ is sufficient to explain the Tibet AS$\gamma$ flux, where $d$ is the average source distance. Although the mechanism of PeV electron acceleration remains an open question, leptonic sources with such hard spectra and high $E_{e,\rm max}$ have been previously observed~\citep[see, e.g.,][]{2018arXiv181001892H,2019arXiv190908609H, 2020ApJ...889L...5F}. 

The Galactic diffuse emission at GeV-TeV energies is expected to be contributed by the $\pi^0$ decay of hadronic cosmic rays and the inverse-Compton emission by diffuse electrons and electrons from pulsars  \citep{Ackermann_2012,PhysRevLett.120.121101}. 
The dash-dotted curve in Figure~\ref{fig:Leptonic} shows such a component. It shows one of our benchmark case (cosmic-ray model I and uniform source distribution) but with 50\% of the best-fit normalization.

\section{Conclusions and Discussion}\label{sec:discussion}
The flux level of Galactic neutrinos has been a mystery and remains undetected by the current-generation neutrino telescopes. Using the recently measured Galactic diffuse gamma-ray intensity, we derived the neutrino flux from the GP. We took into account uncertainties in the gamma-ray attenuation and cosmic-ray spectrum, and the all-sky-averaged neutrino intensity from the GP is estimated to be $\lesssim(3-6)\times10^{-9} \,\rm GeV\,cm^{-2}\,s^{-1}\,sr^{-1}$ at 100~TeV. Our results are consistent with both previous gamma-ray and neutrino constraints. Our calculation relies on the simple connection between neutrino and gamma rays produced by $pp$ interactions, so the results on the sub-PeV neutrino flux are not much sensitive to the chemical composition of cosmic rays as well as details of the propagation and source distribution. 

The diffuse neutrino intensity toward the GP may be comparable to the extragalactic neutrino intensity in the same sky region. Given that the GP is $\sim10$\% of the sky, the detection is promising for next-generation telescopes such as KM3Net and IceCube-Gen2, which may also find a large-scale anisotropy due to the Galactic component \citep{2020PhR...872....1B}. 
The seven-year all-flavor IceCube GP sensitivity around 100~TeV is $\sim 10^{-8}\,\rm GeV\,cm^{-2}\,s^{-1}\,sr^{-1}$. With a factor of $\sim5$ increase in the neutrino effective area \citep{2020arXiv200804323T}, the GP sensitivity of IceCube-Gen2 may reach $\sim3\times10^{-9}\,\rm GeV\,cm^{-2}\,s^{-1}\,sr^{-1}$ in ten years, while KM3Net \citep{AIELLO2019100} would reach $\sim (3-6)\times10^{-9}\,\rm GeV\,cm^{-2}\,s^{-1}\,sr^{-1}$ at 100~TeV for the diffuse Galactic neutrino spectrum, although the detailed value depends on the spectral shape. These would be sufficient to detect the diffuse neutrino flux indicated by the lower bound of the blue shaded region in Figure~\ref{fig:fullSkyAveragedPhi}.

The origin of neutrinos below 100~TeV has emerged as a new mystery~\citep{Aartsen:2015ita,Aartsen:2020aqd}. The measured spectral index ($\sim2.53$) is compatible with that of diffuse GP emission. With these similar indices, the Tibet AS$\gamma$ data imply that the GP contribution remains $\lesssim10$\% even at $\sim10-100$ energies, further supporting the manifestation of extragalactic origins~\citep{Murase:2015xka}. We do not exclude possibilities that a fraction of neutrinos come from other regions such as the Galactic halo~\citep[see Section~2 of][]{Ahlers:2013xia}, although the Tibet AS$\gamma$ off-source data imply a Galactic halo contribution lower than $E_\nu^2\Phi_\nu\lesssim2\times{10}^{-9}~{\rm GeV}~{\rm cm}^{-2}~{\rm s}^{-1}~{\rm sr}^{-1}$. Other Galactic sources, such as the {\it Fermi} bubbles, have also been constrained \citep{Ahlers:2013xia,Lunardini:2013gva,2017PhRvD..96l3007F}. These are consistent with the fact there is no significant northern-southern asymmetry in the neutrino sky~\citep{Aartsen:2020aqd}.    

While the diffuse Galactic interpretation of the Tibet AS$\gamma$ data seems the most natural, discrete sources may still significantly contribute especially at the highest energies. This is especially the case if the cosmic-ray nucleon spectrum is as steep as $E^{-2.7}$ with a break energy of $\lesssim 1$~PeV. If a crucial fraction of the highest-energy events detected by the Tibet AS$\gamma$ experiment is associated with the Cygnus Cocoon, the presence of an efficient Pevaron would be supported. The Tibet AS$\gamma$ data can also be explained by unresolved Pevatrons such as hypernova remnants in the Cygnus region and (or) other part of the Galaxy. Finally, the leptonic scenario is not excluded. Future multi-messenger observations by not only neutrino telescopes but also near-future gamma-ray experiments such as LHAASO, ALPACA, and SWGO are necessary to discriminate between these scenarios. The spatial distribution would give us crucial information, and observations in the southern sky are relevant~\citep{Ahlers:2013xia, 2019BAAS...51g.109H}. A few or dozens of sources are sufficient to explain the sub-PeV gamma-ray intensity, which is promising for source identification. 

\bigskip
While preparing the manuscript, we became aware that \citet{2021arXiv210402838D, 2021arXiv210403729Q,2021arXiv210405609L} appeared on arXiv. Our work was carried out independently. 

We thank Markus Ahlers, Julia Becker Tjus, Kazumasa Kawata, and Walter Winter for useful comments and communications. 
The work of K.F. is supported by the Office of the Vice Chancellor for Research and Graduate Education at the University of Wisconsin–Madison with funding from the Wisconsin Alumni Research Foundation. The work of K.M. is supported by NSF Grant No.~AST-1908689, and KAKENHI No.~20H01901 and No.~20H05852. 

\bibliography{references,kmurase}

\begin{appendix}

\section{Optical Depth for Sub-PeV Gamma Rays}\label{Appendix:IR}
\begin{figure*}[htp]
\centering
\includegraphics[width=.49\textwidth]{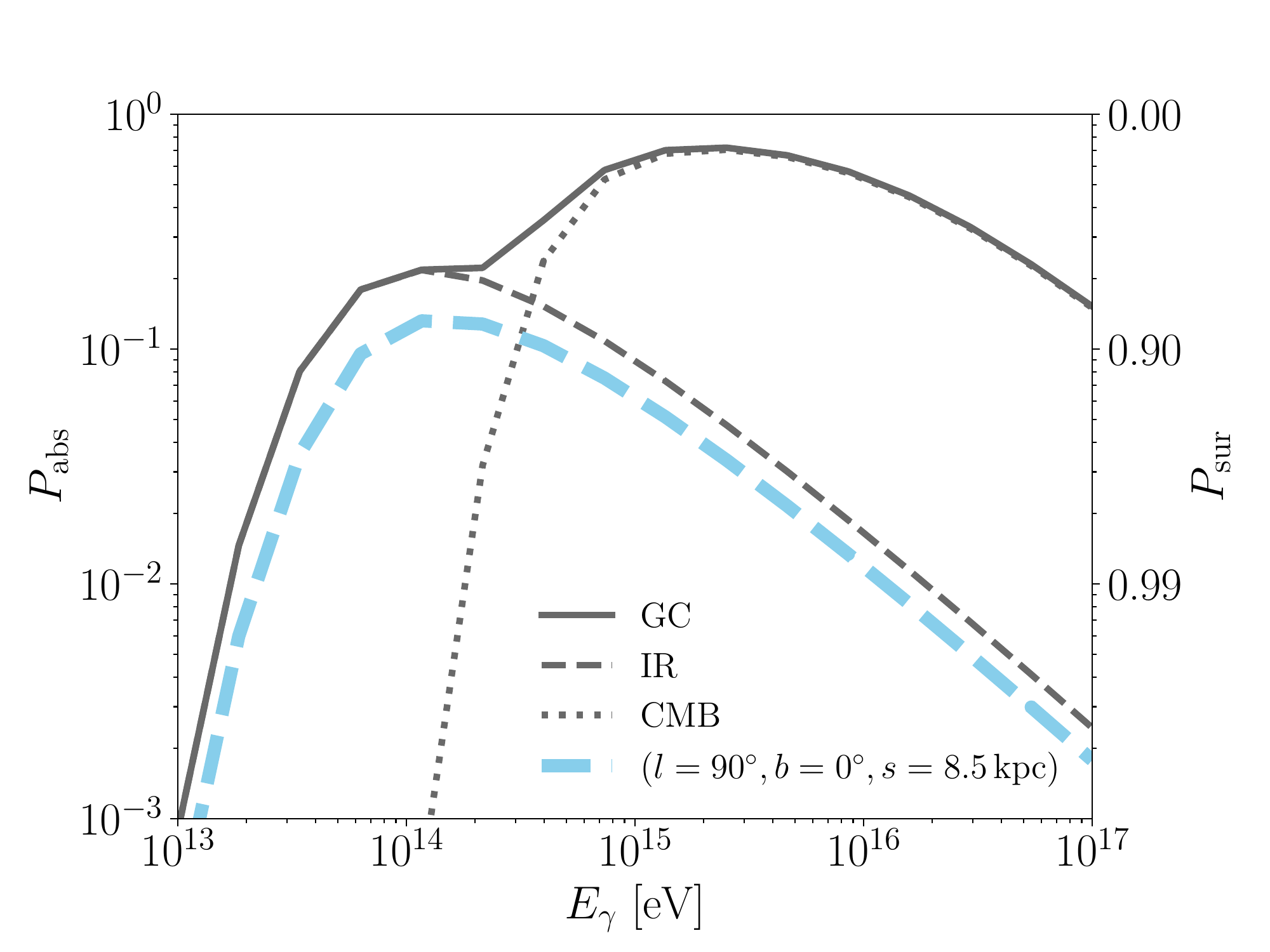}
\includegraphics[width=.49\textwidth]{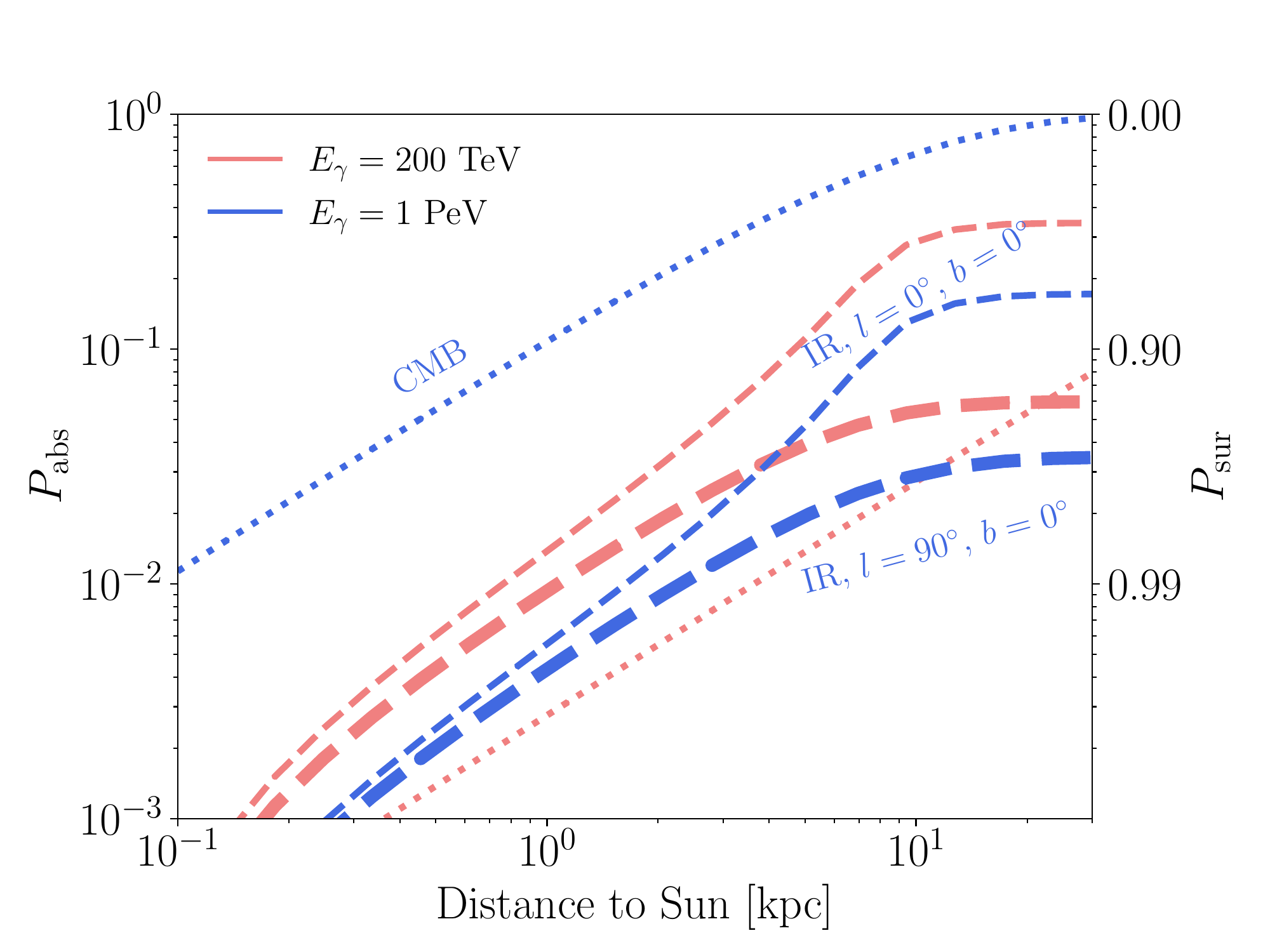}
\caption{\label{fig:Psurv} Survival probability of gamma rays travelling to an observer in the solar neighborhood. The attenuation due to the CMB is indicated by dotted curves and that due to the infrared dust emission is indicated by dashed curves.
{\bf Left:} the survival probability as a function of gamma-ray energy for photons starting from the Galactic Center (GC, grey) and a certain direction ($l=90^\circ,\,b=0^\circ$, $s=8.5\,\rm kpc$) inside the sky region where diffuse gamma rays are observed (light blue). 
{\bf Right:} $P_{\rm surv}$ as a function of distance to the Sun. Gamma-ray energy is indicated in color (red corresponds to $E_\gamma = 200$~TeV and blue corresponds to $E_\gamma = 1$~PeV) and the direction is indicated by the thickness of the curves (thin dashed indicates the GC direction and thick dashed indicates $l=90^\circ, b=0^\circ$).}
\end{figure*}

Gamma rays above a few 100~TeV mostly pair-produce with the cosmic microwave background (CMB). The differential number density of the isotropic black body emission is
\begin{equation}
\left.\frac{dn}{d\varepsilon d\Omega}\right|_{\rm CMB} = \frac{2\varepsilon^2}{h^3c^3}\frac{1}{e^{\varepsilon/(k_B T_{\rm CMB})}-1},
\end{equation}
where $h$ is the Planck constant, $k_B$ is the Boltzmann constant, and $T_{\rm CMB}=2.73$~K is the CMB temperature. 

Gamma rays between $\sim$10~TeV and $\sim 1$~PeV also interact with the dust emission with wavelength $\lambda \gtrsim 50\,\mu$m. We follow \citet{Vernetto_Lipari16} for the calculation of the infrared density field of the Galaxy. The intensity of the infrared emission at the location $\vec{x}$ and direction $\hat{k}$ is obtained by integrating the dust emissivity along the line-of-sight $s$, 
\begin{equation}
    I_\lambda (\vec{x}) = \int_0^\infty ds\, \eta_\lambda(\vec{x} + \hat{k}s).
\end{equation}
Here $\eta_\lambda$ is the power emitted per unit volume, unit solid angle and unit wavelength by the dust, 
\begin{equation}
    \eta_\lambda = \rho_d \kappa_\lambda B_\lambda(T)
\end{equation}
$\kappa_\lambda$ is the absorption cross section per mass of dust, for which we have used the $\kappa_\lambda$ values from \citet{2003ARA&A..41..241D} (the $R_V=3.1$ model).  $B_\lambda(T)$ is the spectral radiance of a black body,
\begin{equation}
    B_\lambda(T) = \frac{2hc^2}{\lambda^5}\frac{1}{e^{hc / (\lambda k_B T)}-1}. 
\end{equation}
The infrared emission comes from cold and warm dust components. Their density and temperature profiles are assumed to follow the disk structure, 
\begin{equation}
    \rho_{c, w}\propto \exp{\left[-\frac{r}{r^0_{c,w}} - \frac{|z|}{z^0_{c, w}}\right]}.
\end{equation}
We adopt the $\rho^0_{c, w}$ and $T^0_{c, w}$ expressions described by equations~16 and 17 of \citet{Vernetto_Lipari16}.  
The number density of the dust emission is computed by 
\begin{equation}
     \left.\frac{dn}{d\varepsilon d\Omega}\right|_{\rm IR} (\vec{x})= \frac{1}{c \,\varepsilon}  I_\varepsilon(\vec{x})
\end{equation}
where $I_\varepsilon = I_\lambda \lambda / \varepsilon$. 
The extragalactic background light (EBL) may also interact with gamma rays. However, it is sub-dominant comparing to the Galactic dust emission and therefore ignored in our calculation. 

Figure~\ref{fig:Psurv} compares the attenuation effect of CMB and infrared photons for gamma rays traveling from the different directions at various energies. Gamma rays above $\sim$500~TeV are mostly absorbed by the CMB. Gamma rays between $\sim 50$~TeV and $\sim 500$~TeV may pair produce with the dust emission. The survival probability depends on the direction and distance of the gamma-ray source. Due to the spatial distribution of the dust, gamma rays from the inner Galaxy is more absorbed than those from the other parts of the Galaxy.

\section{Diffuse Galactic Neutrino Spectrum}\label{appendix:nuSpectra}
\begin{figure}[htp]
\centering
\includegraphics[width=.49\textwidth]{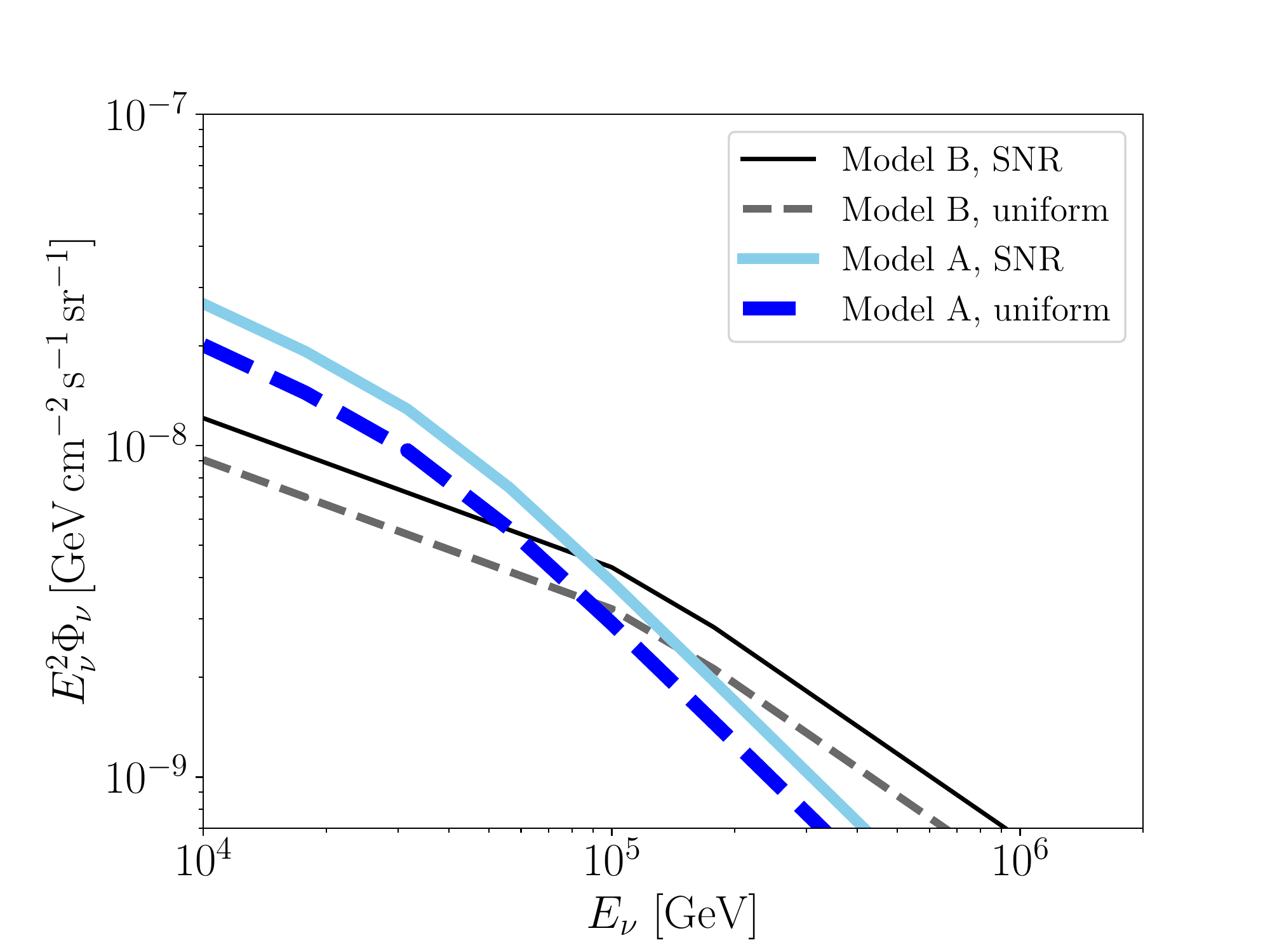}
\caption{\label{fig:fullSkyAveragedPhiModels} 
{All-sky-averaged intensities of all flavor diffuse neutrinos from the GP, for the two neutrino models described by equations~\ref{eqn:Gaisser} (Model A) and \ref{eqn:break} (Model B) and the two source distribution models described by equations~\ref{eqn:uniform} (uniform) and \ref{eqn:SNR} (SNR). The lowest and highest fluxes in each energy bin are used as the boundary of the blue filled area in Figure~\ref{fig:fullSkyAveragedPhi}. The intensity normalization is set by the Tibet AS$\gamma$ data.}
}
\end{figure}

The cosmic-ray nucleon spectrum between 10~TeV and 10~PeV has been studied by different groups based on the measured cosmic-ray spectrum and modeling of the chemical composition~\citep[e.g.,][]{2003APh....19..193H,Gaisser:2012zz,2013EPJWC..5209004G, 2016A&A...595A..33T, 2020APh...12002441L}. The fitting to observation relies on the assumption of the spectral model, including the dependence of the break energy on the charge and the mass number of each chemical group. It also depends on the choice of hadronic interaction models. For example, based on direct measurements of cosmic rays, \cite{2003APh....19..193H} presented the polygonato model with the rigidity-dependence hypothesis, where the rigidity cutoff is set to 4~PeV~\citep[see also][]{Gaisser:2012zz}. However, the proton spectral break may exist well below the knee energy, and nuclei such as helium may be dominant around the knee~\citep[see also][]{Gaisser:2013bla,2020APh...12002441L}. More recently, \citet{2020APh...12002441L} showed that different proton spectral indices are needed to explain the KASCADE \citep{2013arXiv1306.6283T} and IceTop/IceCube data \citep{2019PhRvD.100h2002A} that are interpreted with QGSJet and Sibyll models. 
To account for uncertainties in the cosmic-ray spectrum and composition, we consider two models for the diffuse Galactic neutrino spectrum. In Model A, following \cite{Ahlers:2013xia}, we adopt
\begin{equation}\label{eqn:Gaisser}
\frac{dN}{dE_\nu} \propto E_\nu^{-\alpha_{\nu}}\left[1+\left(\frac{E_\nu}{E^*}\right)^w\right]^{-\delta/w} 
\end{equation}
with $\alpha_\nu=2.54$, $\delta=0.67$, $E_{\nu,*}=0.04\times0.9$~PeV and $w=3.0$. The spectral index is based on the nucleon spectrum obtained by \citet{2013EPJWC..5209004G} with a $0.1$ spectral hardening to account for the energy dependence of the inelastic $pp$ cross section \citep[e.g.,][]{2006ApJ...647..692K,Kelner:2006tc}. 
The nucleon spectral cutoff, 0.9~PeV, is motivated by the limit that cosmic rays around the knee energy are dominated by helium nuclei with $A=4$~\citep{Gaisser:2012zz}.
The factor of $\sim0.04$ converts nucleon energy to neutrino energy, since leading neutrinos from $pp$ interactions carry $\sim(3-4)\%$ of the nucleon energy in this energy range. 

In Model B, the neutrino spectrum is assumed to be a broken power law,  
\begin{equation}\label{eqn:break}
\frac{dN}{dE_\nu} \propto 
\begin{cases} E_\nu^{-\alpha_{\nu1}} & E_\nu < E_{\nu, \rm bk}  \\ 
E_\nu^{-\alpha_{\nu2}} & E_\nu > E_{\nu, \rm bk}
\end{cases}
\end{equation}
where $\alpha_{\nu1}=2.45$, $\alpha_{\nu2}=2.85$, and $E_{\nu,\rm bk}=0.04\times3$~PeV. This model is based on modeling of the all-particle cosmic-ray spectrum with a nucleon spectral cutoff energy of $\sim3-4$~PeV comparable to the knee energy~\citep[e.g.,][]{2003APh....19..193H,Gaisser:2012zz,2020APh...12002441L}, but we further take possible hardening of the cosmic-ray spectral index due to the spatial inhomogeneity~\citep[by 0.15;][]{Acero:2016qlg,Lipari:2018gzn}. The spectrum around the knee is believed to be largely contributed by helium nuclei, so the break energy is likely to be lower. This choice of the break energy and spectral index should be regarded as the most optimistic case. 

\end{appendix}

\end{document}